\documentclass[
  journal=pasa,
  manuscript=research-paper,
  year=2020,
  volume=37,
]{cup-journal}
\usepackage{hyperref}
\usepackage{amsmath}
\usepackage{booktabs}

\title{Bayesian evidence-driven likelihood selection for sky-averaged 21-cm signal extraction}

\author{K. H. Scheutwinkel}
\affiliation{Astrophysics Group, Cavendish Laboratory, J. J. Thomson Avenue, Cambridge CB3 0HE, UK}
\alsoaffiliation{Kavli Institute for Cosmology, Madingley Road, Cambridge CB3 0HA, UK}
\email[K. H. Scheutwinkel]{khs40@cam.ac.uk}

\author{W. Handley}
\affiliation{Astrophysics Group, Cavendish Laboratory, J. J. Thomson Avenue, Cambridge CB3 0HE, UK}
\alsoaffiliation{Kavli Institute for Cosmology, Madingley Road, Cambridge CB3 0HA, UK}

\author{E. de Lera Acedo}
\affiliation{Astrophysics Group, Cavendish Laboratory, J. J. Thomson Avenue, Cambridge CB3 0HE, UK}
\alsoaffiliation{Kavli Institute for Cosmology, Madingley Road, Cambridge CB3 0HA, UK}


\begin{document}

\begin{abstract}
We demonstrate that the Bayesian evidence can be used to find a good approximation of the ground truth likelihood function of a dataset, a goal of the likelihood-free inference (LFI) paradigm. As a concrete example, we use forward modelled sky-averaged 21-cm signal antenna temperature datasets where we artificially inject noise structures of various physically motivated forms. We find that the Gaussian likelihood performs poorly when the noise distribution deviates from the Gaussian case e.g. heteroscedastic radiometric or heavy-tailed noise. For these non-Gaussian noise structures, we show that the generalised normal likelihood is on a similar Bayesian evidence scale with comparable sky-averaged 21-cm signal recovery as the ground truth likelihood function of our injected noise. We therefore propose the generalised normal likelihood function as a good approximation of the true likelihood function if the noise structure is a priori unknown.
\end{abstract}

\section{INTRODUCTION}

21-cm cosmology \citep{furlanetto_cosmology_2006} studies the hydrogen hyperfine transition line with a characteristic free-space wavelength of $\lambda = 21$ cm. When this spin-flip transition is emitted during Cosmic Dawn (CD) and Epoch of Reionisation (EOR), the wavelength is redshifted to lower frequencies, enabling us to probe these epochs through radio observatories. Experimental approaches such as the sky-averaged experiment or also known as the global experiment \citep{liu_global_2013} measure the contrast between integrated absorption and emission of the 21-cm signal and the cosmic microwave background (CMB) across the whole sky. This contrast which is known as the sky-averaged 21-cm signal has a characteristic shape \citep{pritchard_evolution_2008} that can be probed to study and constrain the earliest epochs of the universe. 

The EDGES collaboration \citep{bowman_absorption_2018} were the first to report a deep flattened absorption feature at CD with an amplitude of $500^{+500}_{-200}$ mK centred at $78\pm 1$ MHz. However, these results are not compatible e.g. with the astrophysical models of \cite{cohen_charting_2017} which predicted a weaker absorption signal with less than a half of the amplitude of what the EDGES collaboration reported. Furthermore, the best-fitting profile reported by EDGES is not a unique solution, however, the most physically justifiable one and an unmodelled sinusoidal systematic structure can be found when changing the foreground model i.e. the number of polynomials  \citep{hills_concerns_2018}. These discrepancies caused an open scientific debate \citep{bowman_reply_2018} that can only be resolved by an independent sky-averaged experiment such as SARAS 2 and 3 \citep{singh_saras_2018-1, singh_detection_2021}, LEDA \citep{price_design_2018}, REACH \citep{de_lera_acedo_reach_2022} or many others. Possible candidates to explain the EDGES signal, if of cosmological origin, could involve new physics; an excessive cooling mechanism of the IGM through dark matter particle interaction \citep{barkana_signs_2018, barkana_possible_2018, munoz_insights_2018} or an enhanced radio background apart from the CMB \citep{jana_radio_2019, fialkov_signature_2019, mirocha_what_2019} with standard stellar population models incapable to achieve this \citep{mittal_implications_2022}. However, some argue that the claim of new physics is not justified as this deep absorption feature could be a result of unmodelled systematic features such as a ground plane artefact \citep{bradley_ground_2019}, calibration issues \citep{sims_testing_2020} or within the data analysis methods e.g. by changing the polynomial foreground model to maximally smooth functions \citep{bevins_maxsmooth_2021, singh_redshifted_2019}. Additionally, recent data collected by SARAS 3 \citep{singh_detection_2021} reject the best-fitting profile of EDGES with 95.3\% confidence.

Adding these various model components into the simulations and comparing them with the observations is the subject of current state-of-the-art sky-averaged 21-cm cosmological research. However, determining the statistical properties of this often hierarchical generative model which potentially consists of many latent hyperparameters poses a challenge for current Bayesian inference tasks. This is due to the necessity of an explicit likelihood expression to recover the parameters of our model i.e. solving the inverse problem of the sky-averaged 21-cm simulations. Oversimplification of the noise structure e.g. through a Gaussian approximation can engender inaccurate posterior inferences.

We investigate this problem of unknown noise structures, by generating antenna temperature datasets with non-Gaussian or heavy-tailed noise and study its influence on the sky-averaged 21-cm signal parameter inference by using likelihood functions of various forms. We use a Bayesian evidence-driven likelihood selection to decide which likelihood is preferred and the best approximation of the (unknown) ground truth likelihood of our dataset. This Bayesian evidence-driven likelihood selection analysis can be embedded into the Likelihood-Free Inference (LFI) paradigm \cite{marin_approximate_2011, papamakarios_fast_2016, cranmer_frontier_2020, papamakarios_normalizing_2021}, where the goal is to find a satisfactory approximation of the unknown ground truth likelihood function.
Examples of various algorithmic LFI methods applied in 21-cm cosmology can be found in \cite{zhao_implicit_2022, zhao_simulation-based_2022} or in the broader cosmological field in \cite{alsing_massive_2018, alsing_fast_2019, jeffrey_likelihood-free_2020}.

In Section \ref{sec:BayesianInf}, we briefly explain the Bayesian Inference framework and its parameter estimation and model comparison component. In Section \ref{sec:ForwardModel}, we describe how we generate the sky-averaged 21-cm signal antenna temperature datasets using a physically motivated forward model or also known as a simulator. In Section \ref{sec:BayesModel}, we present various candidates of likelihood functions to statistically model our (unknown) datasets and in Section \ref{sec:Results},  we present our findings. Finally, we summarise our work in Section \ref{sec:Summary}.

\section{BAYESIAN INFERENCE AND NESTED SAMPLING}
\label{sec:BayesianInf}
We use the Bayesian inference framework \citep{sivia_data_2006} to analyse a dataset $D$, which is a powerful statistical framework based on Bayes Theorem of conditioned probabilities:
\begin{equation}
    \mathcal{P} = \frac{\mathcal{L} \pi}{\mathcal{Z}},
\end{equation}
where $\mathcal{P} \equiv p(\theta|D, M)$ is the posterior distribution of the parameters $\theta$ of model $M$ after the dataset $D$ has been observed. The posterior is recovered by combining the likelihood  $\mathcal{L} \equiv p(D|\theta, M)$ of the dataset with the prior $\pi \equiv p(\theta|M)$ of the parameters, a distribution containing prior assumptions before any data has been observed. The product of these two distributions is normalised by $\mathcal{Z}$ the marginalised likelihood or also known as the Bayesian evidence.

With the help of algorithms such as Markov-Chain Monte Carlo (MCMC) methods \citep{mackay_information_2003} one is able to address the parameter estimation problem of Bayesian inference to generate posterior samples $\theta^*$ from the prior. However, MCMC methods avoid solving the generally computationally expensive integration of the marginalised likelihood:
\begin{equation}
    \mathcal{Z} = \int \mathcal{L}(\theta) \pi(\theta) d\theta,
\end{equation} which is an integration over all possible parameter values $\theta$ of the model $M$. The solution of the integral is the Bayesian evidence $\mathcal{Z}$, a key component for model comparison in Bayesian inference. As the model comparison is essential for this analysis, we use a nested sampling-based \citep{skilling_nested_2006} algorithm which numerically computes the Bayesian evidence. We choose \texttt{PolyChord} \citep{handley_polychord_2015-1, handley_polychord_2015} as our nested sampler, which utilises a slice sampling method to generate new proposal candidates subject to an evolving likelihood constraint. Additionally, \texttt{PolyChord} proves to be a more scalable solution than \texttt{MultiNest} \citep{feroz_multimodal_2008, feroz_multinest_2009} which is vital for high-dimensional parameter spaces. Moreover, as \texttt{PolyChord} is a sampling-based method, we can acquire posterior samples and, therefore, address the model comparison and parameter estimation part of Bayesian
inference simultaneously. 

With the Bayesian evidence $Z \equiv p(D|M)$ we recover the model probabilities by applying Bayes Theorem:
\begin{equation}
    p(M|D) = \frac{p(D|M)p(M)}{p(D)}.
\end{equation}
 For two competing models $M_1$ and $M_2$ which are assumed a priori equally likely $p(M_1) = p(M_2)$, we define the Bayes factor:
 \begin{equation}
 \label{eqn:BayesFac}
    \log \mathcal{K} = \log p(M_1|D) - \log p(M_2|D),
\end{equation}
which is a Bayesian evidence posterior ratio of both models given our assumption. Hence, a positive Bayes factor indicates the preference of model $M_1$ for the dataset $D$.

\section{DATASET SIMULATION}
\label{sec:ForwardModel}
To generate the dataset $D$, we use a physically motivated simulator. The simulator splits the dataset into a sum of three components: 
\begin{equation}
\label{eqn:dataset}
    D\equiv T_{\mathrm{data}} = T_{\mathrm{fg}} + T_{21} + T_{\mathrm{noise}},
\end{equation}
where $T_{\mathrm{fg}}$ is the foreground component, $T_{21}$ the sky-averaged 21-cm signal and $T_{\mathrm{noise}}$ the contribution of the noise model. We describe in the subsequent sections how we simulate each component.

\subsection{Foreground simulation}
To simulate the foreground we use the Bayesian foreground modeling framework of \cite{anstey_general_2021} or also referred to as the Bayesian data analysis pipeline with its standard settings that are being used in REACH \citep{de_lera_acedo_reach_2022} e.g. to quantify the influence of unmodelled systematic structure for sky-averaged 21-cm signal parameter inference \citep{scheutwinkel_bayesian_2022}. The pipeline uses the Global Sky Model (GSM) map \citep{de_oliveira-costa_model_2008} to construct two maps at 408 MHz and 230 MHz frequencies. With these maps, we construct the spatially varying spectral index parameter:
\begin{equation}
\label{eqn:beta}
    \beta(\Omega) = \frac{\log \left(\frac{T_{230}(\Omega) - T_{\mathrm
    {CMB}}}{T_{408}(\Omega) - T_{\mathrm
    {CMB}}}\right)}{\log \left(\frac{230}{408} \right)},
\end{equation}
with $T_{\mathrm{CMB}} = 2.375$ K the CMB.  We choose this spatially dependent spectral index as it is physically more realistic than a constant spectral index. Moreover, \cite{anstey_general_2021} investigated the influence of a uniform spectral index for the resulting sky-averaged 21-cm parameter inference and concluded that a uniform index introduces spectral features which are mimicking a sky-averaged 21-cm signal, hence, making the signal extraction unnecessarily difficult or unsuccessful.
We use this spectral index $\beta(\Omega)$ to generate our foreground:
\begin{equation}
    T_{\mathrm{sim}}(\nu, \Omega) = (T_{230}(\Omega) - T_{\mathrm{CMB}} ) \left(\frac{\nu}{230} \right)^{-\beta(\Omega)} + T_{\mathrm{CMB}},
\end{equation}
where we use $T_{230}(\Omega)$ as a base map such that we generate an approximate representation of the sky in the 50-200 MHz frequency band where the sky-averaged 21-cm signal is expected. Furthermore, 230 MHz is ``far enough'' that there is negligible sky-averaged 21-cm signal contamination to be expected, therefore mitigating potential issues when we add a simulated absorption feature later on. We convolve the foreground map with the conical log-spiral antenna beam pattern $B_{\Omega}$ \citep{dyson_characteristics_1965}:
\begin{equation}
    T_{\mathrm{fg}}(\nu) =\frac{1}{4 \pi} \int_{\Omega} B(\Omega,\nu) T_{\mathrm{sim}}(\Omega,\nu) d\Omega
\end{equation} to compute the resulting antenna temperature $T_{\mathrm{fg}}$. 
The log-spiral beam pattern is reported to have the most accurate capabilities to recover the sky-averaged 21-cm signal compared to five other antenna designs \citep{anstey_informing_2021}. The antenna beam pattern is FEKO simulated \citep{elsherbeni_antenna_2014}, assumed to be in ideal conditions using an infinite ground plane to mitigate edge effects. This setup is located at $\lambda = -30.71131$° and $\phi = 21.4476236$° at the Karoo Radio reserve in South Africa, where REACH is currently being deployed. We choose a snapshot of the sky at `2019-10-01 00:00:00' UTC (LST 23.99 h), where the Milky Way centre is not at the zenith to mitigate brighter foreground contributions.
\subsection{Sky-averaged 21-cm signal}
To simulate the sky-averaged 21-cm signal, we add a Gaussian absorption profile which is a Gaussian approximation of the standard case of the astrophysical models of \cite{cohen_charting_2017}:
\begin{equation}
\label{eqn:G21Signal}
    T_{\mathrm{21}}(\nu) = - A_{21} \exp \left(- \frac{1}{2} \left(\frac{\nu - f_{0,21}}{\sigma_{21}} \right)^2 \right),
\end{equation}
where $A_{21}$ is the amplitude of the absorption signal, $\sigma_{21}$ the standard deviation, $f_{0,21}$ the central frequency and $\nu$ the frequency. For our absorption feature, we set $A_{21} = 155$ mK, $\sigma_{21} = 15$ MHz and $f_{0,21} = 85$ MHz.
\subsection{Antenna temperature noise}

\begin{figure}
    \centering
    \includegraphics{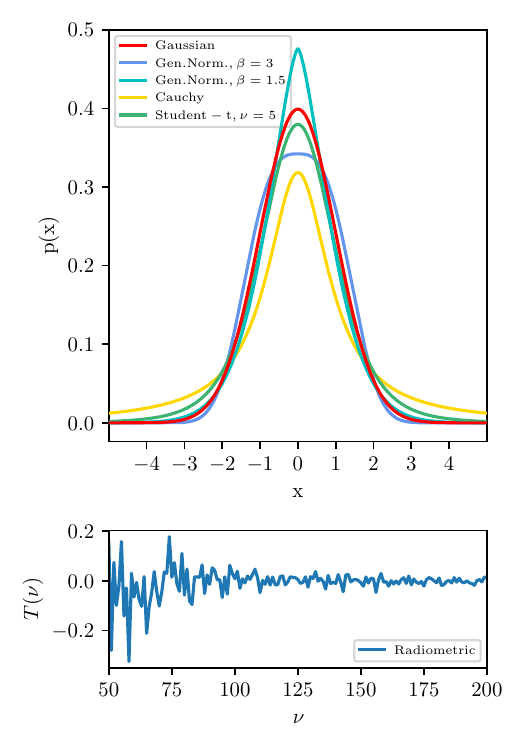}
    \caption{Top: Comparison of standardised probability density functions. Bottom: Exemplary heteroscedastic radiometric noise realisation that decreases exponentially $\propto \nu^{-2.6}$ over a range $\nu$.}
    \label{fig:noisepdfs}
\end{figure}

The Bayesian data analysis pipeline of \cite{anstey_general_2021} has demonstrated the success of the sky-averaged 21-cm signal recovery through simulating antenna temperature dataset with homoscedastic Gaussian noise. However, this noise structure is not an accurate physically motivated choice, hence, we add varying non-Gaussian noise models onto the antenna temperature $T_{\mathrm{fg}}$ to study its influence on the sky-averaged 21-cm recovery. As non-Gaussian noise can potentially arise through imperfections in a real experimental setup or through complex unknown noise structures within a physical process, we consider one of the following noise models for each dataset which are also visualised in Figure (\ref{fig:noisepdfs}).

\begin{enumerate}
    \item Gaussian noise:
\begin{equation}
    p(x; \mu, \sigma) = \frac{1}{\sqrt{2\pi \sigma^2}} \exp \left(-\frac{(x-\mu)^2}{2\sigma^2}\right)\,,
\end{equation}
with mean $\mu$ and standard deviation $\sigma$.
\item 
Generalised normal noise:
\begin{equation}
    p(x; \mu, \alpha, \beta) = \frac{\beta}{2 \alpha \Gamma (1/ \beta)} \exp \left(- (|x-\mu|/\alpha)^\beta \right)\,,
\end{equation}
with mean $\mu$, shape parameter $\beta$ and scale parameter $\alpha$. If $\beta = 2$, this noise model recovers the Gaussian noise model. Furthermore, the variance of the distribution can be expressed in terms of the shape and scale parameter: $\sigma^2 = \frac{\alpha^2 \Gamma(3/\beta)}{\Gamma(1/\beta)}$.
\item 
The Student-t distribution:
\begin{equation}
    p(x; \nu, \hat{\mu}, \hat{\sigma}) = \frac{\Gamma (\frac{\nu + 1 }{2})}{\sqrt{\nu \pi} \hat{\sigma} \Gamma (\frac{\nu}{2})}\left(1 + \frac{1}{\nu} \left( \frac{x - \hat{\mu}}{\hat{\sigma}} \right)^2 \right)^{- \frac{\nu + 1}{2}}\,,
\end{equation}
with $\nu$ degrees of freedom, the location parameter $\hat{\mu}$ and the scale parameter $\hat{\sigma}$. This distribution arises after the antenna temperature has been calibrated in a Bayesian way \citep{roque_bayesian_2021}.
\item The Cauchy distribution:
\begin{equation}
    p(x; x_0, \gamma) = \frac{1}{\pi \gamma \left[1 + \left( \frac{x-x_0}{\gamma} \right)^2 \right]}\,,
\end{equation}
with the location parameter $x_0$ and the scale parameter $\gamma$. This distribution has an undefined mean and variance, therefore, it is a heavy-tail distribution simulating frequent outliers i.e. extremely noisy structures. 
\item The radiometric noise \citep{kraus_radio_1986},
a physically motivated radiometer model, which is defined as:
\begin{equation} \label{eqn:radiometricN}
    \sigma_{\mathrm{radio}}(\nu) = \frac{\eta T_{\mathrm{fg}}(\nu) + (1-\eta) T_0 + T_{\mathrm{rec}}}{\sqrt{\Delta \nu \tau}}\,,
\end{equation}
where $T_{\mathrm{fg}}$ is the antenna temperature, $\eta$ the antenna radiation efficiency, $T_{\mathrm{rec}}$ the antenna receiver temperature, $\tau$ the integration time and $\Delta \nu$ the channel width. We model the noise through a Gaussian distribution with heteroscedastic frequency-dependent radiometric noise.
\end{enumerate}

For each noise model, we centre the probabilistic distribution at the mean/location parameter $\mu = T_{\mathrm{fg}}$, set the standard deviation/scale parameter to $\sigma = 0.025$ K and vary the shape parameter if a shape parameter exists, otherwise, we vary the scale parameter. The scale parameter represents values seen in EDGES after removing the foreground and 21-cm models which are also used as default settings in the REACH data analysis pipeline. For the radiometric noise model, we assume a realistic choice of parameters with $\eta = 0.9$, $T_0 = 293.15$ K, $T_{\mathrm{rec}} = 500$ K, $\Delta \nu = 0.1$ MHz and a variable integration time parameter $\tau$ that stays constant across the frequency band i.e. we assume there is no simulated RFI contamination that requires certain frequency channels to be time-integrated differently.

\section{BAYESIAN MODELLING OF DATASET}
\label{sec:BayesModel}
After dataset creation $D$ with varying noise $T_{\mathrm{noise}}$, we define two models:
\begin{equation}
    M_1 \equiv T_{M_1} = T_{\mathrm{fg}} + T_{21},
\end{equation}
the \textit{signal model} with $T_{\mathrm{fg}}$ the foreground component and $T_{21}$ the sky-averaged 21-cm signal component, and 
\begin{equation}
    M_2 \equiv T_{M_2} = T_{\mathrm{fg}},
\end{equation}
the \textit{no-signal model}, where we leave the sky-averaged 21-cm signal unmodelled. In both models, the noise contribution $T_{\mathrm{noise}}$ is modelled through its likelihood function.

The foreground component $T_{\mathrm{fg}}$ utilises the Bayesian foreground modelling framework of \cite{anstey_general_2021}, where the sky is split into $N_{\mathrm{reg}} = 14$ regions with each region having its own uniform index to model the spatially dependent spectral index of the whole sky to construct chromaticity functions. Hence, the foreground parameterises through their spectral indices $\theta_{\mathrm{fg}} = (\beta_{\mathrm{fg},1:N_{\mathrm{reg}}})$. Additionally, we assume that the simulated experiment is located on an infinite ground plane. Hence, we use the same antenna beam pattern for the inference that has been used for simulating the dataset to facilitate computation. Adding a physically realistic ground plane or soil into the simulations can be studied in subsequent research. We expect that a finite ground plane will introduce systematic effects that are handled separately from the noise modelling within REACH \cite{scheutwinkel_bayesian_2022} while introducing soil will be seen as a smooth loss profile of the radiation efficiency \citep{acedo_skala_2015} that can be modelled through the likelihood function.

The precise number of sky regions $N_{\mathrm{reg}}$ is arbitrarily chosen, however, there is a tendency that the simulated datasets with Gaussian noise prefer more than ten regions and the Bayesian evidence plateaus with some fluctuations for higher number of regions \citep{anstey_general_2021}. One can treat this parameter as an extra dimension to ``tune'' the Bayesian evidence to find the optimum sky region number, however, this would increase the number of parameter dimensions of the cosmological model one has to track. Therefore, we do not progress with this extra dimension as the inference is computationally expensive and it has no significant effect on the sky-averaged 21-cm parameter inference as the posterior distributions are seen to be uncorrelated later on. We note that for the analysis of a real experimental dataset, this extra dimension has to be taken into account.

For the sky-averaged 21-cm signal component, we parameterise the Gaussian signal model of eq. (\ref{eqn:G21Signal}) with $\theta_{21} = (f_{0,21}, \sigma_{21}, A_{21})$. For the noise contribution, we have varying parameterisation $\theta_{\mathrm{noise}}$ depending on the likelihood functions used. Therefore, we have the following parameterisation for each model: $\theta_{M_1} =  (\theta_{\mathrm{fg}}, \theta_{21}, \theta_{\mathrm{noise}})$ and  $\theta_{M_2} =  (\theta_{\mathrm{fg}}, \theta_{\mathrm{noise}})$.

\subsection{Likelihood functions}
For each cosmological model $M$, we construct the likelihood functions with:
\begin{equation}
    \label{eqn:GeneralLikelihood}
    \log \mathcal{L} = \log p(T_{\mathrm{data}}|\theta_M, M) =\sum_{\nu} \log  p(T_{\nu}| \theta_M, M),
\end{equation}
which is a sum of the noise model components $p(T_{\nu}| \theta_M, M)$ over the frequencies $\nu$ with $\theta_M$ the parameters of the model $M$ and $T_{\nu}$ the observed antenna temperature. Depending on the choice of the likelihood function, we parameterise the likelihood in the following ways:
\begin{itemize}
    \item Gaussian: $ \theta_{\mathrm{noise}} = \sigma_{\mathrm{L}}$
    \item  Generalised normal: $\theta_{\mathrm{noise}} = (\beta_{\mathrm{L}}, \sigma_{\mathrm{L}})$
    \item Cauchy: $\theta_{\mathrm{noise}} = \gamma_{\mathrm{L}}$
    \item Radiometric: $\theta_{\mathrm{noise}} = (T_{\mathrm{rec}}, \eta, \sigma_{\mathrm{radio}})$

\end{itemize}

For the radiometric likelihood, we define $\sigma_{\mathrm{radio}} = 1 / \sqrt{\tau \Delta \nu}$ as the radiometric noise level. Analogous to the radiometric noise, we model its likelihood function through a Gaussian likelihood with the radiometric noise of eq. (\ref{eqn:radiometricN}) inserted. Moreover, we model the Student-t noise through the generalised normal likelihood as they are similar in nature.

\subsection{Priors}
We list the priors and their ranges in Table \ref{tab:Priors}. We note that for a Bayesian analysis-focused paper, the resulting analysis and result are prior dependent. We choose (log)-uniform priors for all the model parameters as this choice maximises the entropy and puts equal weight across all possible posterior models. One could argue that other prior distributions are more suitable if one understands the underlying system well enough, however, as we do not know what could be a reasonable prior for this problem, we continue our analysis with uniform priors as an enforcement of the principle of maximum entropy. The prior ranges are then set to values that seem the most physically realistic for our experimental setup e.g. setting the foreground region parameters $\beta_{1:N_{\mathrm{reg}}}$ to minimum and maximum values that were explicitly calculated in eq. (\ref{eqn:beta}).

\subsection{\texttt{PolyChord} initialisation}
As $\texttt{PolyChord}$ uses slice sampling within the nested sampling algorithm to draw proposal candidates, $\texttt{PolyChord}$ has sampling initialisations that we list in Table \ref{tab:PolyChordSettings}. Additionally, we run \texttt{PolyChord} twice for each model $M$ and dataset $D$. In the second run, we use the posterior sample mean and sample variance for the foreground region parameters $\beta_{1:N_{\mathrm{reg}}}$of the first iteration to construct narrower prior ranges $\bar \theta^* \pm 5\sigma^*$ where we keep the same prior shape as our initial prior distributions. This ensures proper posterior exploration around high posterior mass regions found in the first run when encountering a potential multimodal problem. An evidence correction has been applied afterwards. This method is adapted from and described in more detail in \cite{anstey_general_2021} and \cite{petrosyan_supernest_2022}. 

\begin{table}
    \centering
    \begin{tabular}{lll}
\toprule
{}
Parameter &     Type         &          Range  \\
\midrule
$\beta_{1:N_{\mathrm{reg}}}$ & uniform &  2.458-3.146 \\
$f_{0,21}$              &      uniform &     50-200 MHz\\
$\sigma_{21}$            &      uniform &      10-20 MHz \\
$A_{21}$            &      uniform &     0-0.25 K \\
\toprule
\textbf{Gaussian noise} \\

$\sigma_{\mathrm{L}}$           &  log uniform &  $10^{-4}$-$10^{-1}$ K \\
\toprule
\textbf{Generalised normal noise} \\
$\sigma_{\mathrm{L}}$           &  log uniform &  $10^{-4}$-$10^{-1}$ K \\
$\beta_{\mathrm{L}}$           &  uniform &  0-5 \\
\toprule
\textbf{Cauchy noise} \\

$\gamma$           &  log uniform &  $10^{-4}$-$10^{-1}$ K \\
\toprule
\textbf{Radiometric noise} \\

$T_{\mathrm{rec}}$           &  log uniform &   100-1000 K \\
$\eta$             &      uniform &     0.8-1 \\
$\sigma_{radio}$           &  log uniform &  $10^{-8}$-$10^{-1}$ \\
\bottomrule
\end{tabular}
    \caption{Prior choices for the foreground spectral indices $\beta_{1:N_{\mathrm{reg}}}$, sky-averaged 21-cm signal shape $f_{0,21}$, $\sigma_{21}$, $A_{21}$ and the varying noise parameters.}
    \label{tab:Priors}
\end{table}

\begin{table}
    \centering
    \begin{tabular}{lll}
\toprule
{} 
Parameter &      Settings                    \\
\midrule
\texttt{$n_{\mathrm{live}}$} & $N_{\mathrm{dim}}*25$  \\
\texttt{$n_{\mathrm{prior}}$} & $N_{\mathrm{dim}}*25$  \\
\texttt{$n_{\mathrm{fail}}$} & $N_{\mathrm{dim}}*25$  \\
\texttt{$n_{\mathrm{repeats}}$}       &    $N_{\mathrm{dim}}*5$ \\
\texttt{precision\_criterion}        &      0.001 \\
\texttt{do\_clustering}       &      True  \\
\bottomrule
\end{tabular}

    \caption{\texttt{PolyChord} initialisations with \texttt{$n_{\mathrm{live}}$} the number of live points, \texttt{$n_{\mathrm{prior}}$} the number of randomly drawn prior samples before execution, \texttt{$n_{\mathrm{fail}}$} the failed spawn criterion, and \texttt{$n_{\mathrm{repeats}}$} the number of slice sampling repeats which is proportional to the model dimension $N_{\mathrm{dim}}$. The \texttt{precision\_criterion} is the termination criterion and \texttt{do\_clustering} activates the clustering algorithm within a nested sampling run.}
    \label{tab:PolyChordSettings}
\end{table}
\section{RESULTS}
\label{sec:Results}

\begin{figure*}
    \centering
    \includegraphics{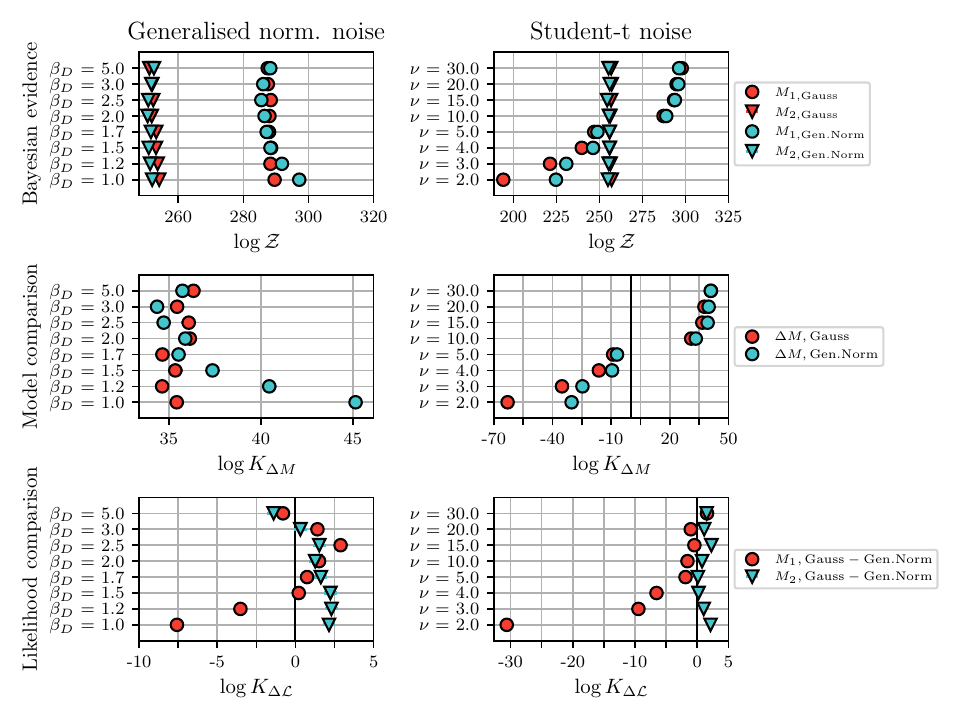}
    \caption{Bayesian evidence $\log \mathcal{Z}$ (first row), signal model comparison ($\Delta M = M_1 - M_2$) (second row) and likelihood comparison Bayes factor (third row) for the Gaussian and generalised normal noise datasets. The errors are in the order of $\sigma_{\log \mathcal{Z}} \approx 0.3$.}
    \label{fig:gauss_gnl}
\end{figure*}

\begin{figure*}
    \centering
    \includegraphics{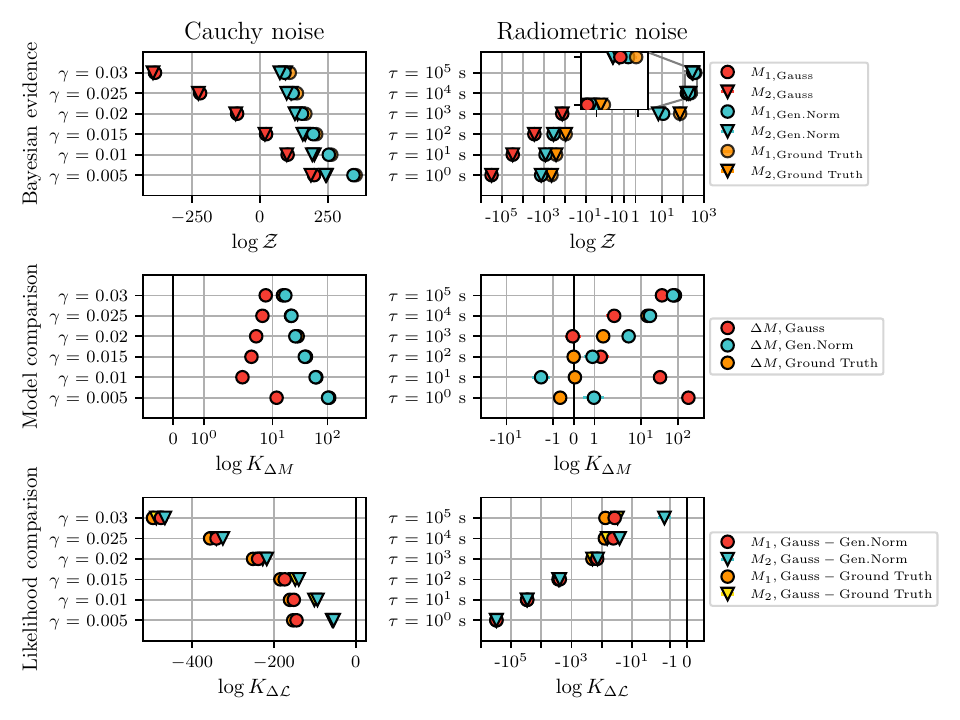}
    \caption{Bayesian evidence $\log \mathcal{Z}$ (first row), signal model comparison ($\Delta M = M_1 - M_2$) (second row) and likelihood comparison Bayes factor (third row)  for the Cauchy and radiometric noise datasets. The errors are in the order of $\sigma_{\log \mathcal{Z}} \approx 0.3$.}
    \label{fig:cauchy_radiom}
\end{figure*}

\begin{figure*}
    \centering
    \includegraphics{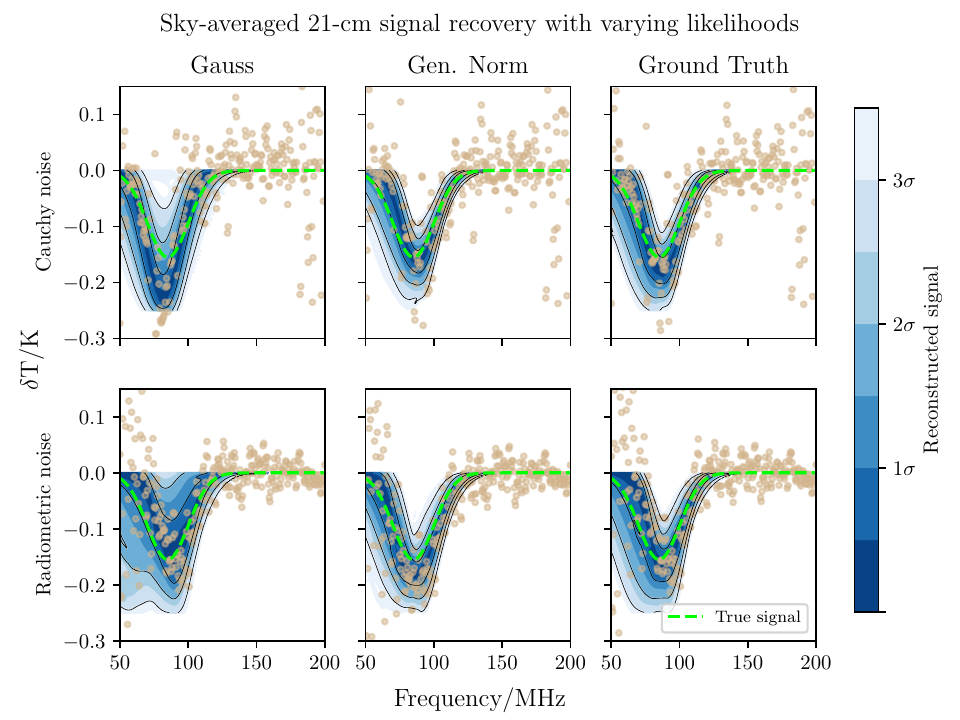}
    \caption{Foreground subtracted residuals (beige) and sky-averaged 21-cm signal recovery (blue) for the Cauchy noise case ($\gamma = 0.025$) (top) and radiometric noise case ($\tau = 10^4$ s) (bottom) when using varying likelihood functions for inference. The upper case shows signs of biased signal mean estimates and the lower case show signs of higher parameter variance.}
    \label{fig:signalrecovery}
\end{figure*}
For each dataset $D$ we use \texttt{PolyChord} to numerically compute the Bayesian evidence $\log \mathcal{Z}$ for the sky-averaged 21-cm signal model $M_1$ and the no signal model $M_2$. In Figures (\ref{fig:gauss_gnl} \& \ref{fig:cauchy_radiom}) we show the corresponding Bayesian evidences and the Bayes factors of the competing signal models for all considered datasets with various injected noise structures. Additionally, for each model $M$, we also vary its likelihood function and present the Bayes factor likelihood comparison. We also provide a Table (\ref{tab:result_evidences}) that lists the computed Bayesian evidences of this study. In Figure (\ref{fig:signalrecovery}), we show exemplary sky-averaged 21-cm signal recoveries when using varying likelihood functions for different noise structures and their implications on the signal parameter inferences.

\subsection{Generalised normal noise}

\begin{figure}
    \centering
    \includegraphics{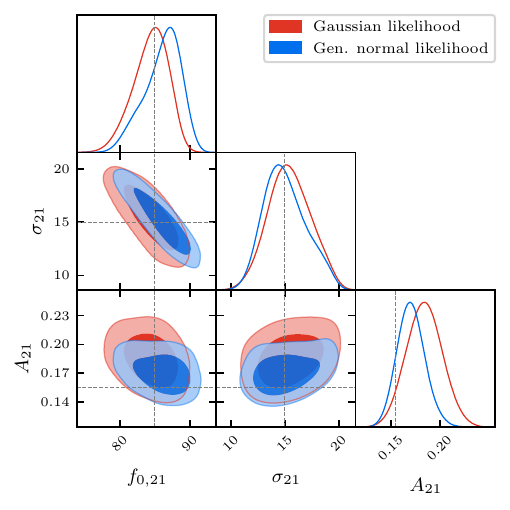}
    \caption{Marginalised posterior estimates of the sky-averaged 21-cm signal parameters when using a Gaussian likelihood (red) versus a generalised normal likelihood function (blue) for an antenna temperature dataset containing generalised normal noise with $\beta_D=1$. The black dashed lines represent the ground truth values of the parameters.}
    \label{fig:triangle_plot_beta1}
\end{figure}
For the generalised normal noise datasets, we note that the signal model $M_1$ has a higher Bayesian evidence than the no signal model $M_2$ irrespective of the noise shape parameter $\beta_D$ and their likelihood functions. This indicates that the sky-averaged 21-cm signal model, the ground truth model of the dataset, is always preferred over the no-signal model with Bayes factors $\log K_{\Delta M} > 34$. However, we see a change in likelihood preference once we change the shape parameter $\beta_D$ of the dataset.

For the case $\beta_{\mathrm{D}}=2$, we observe that the Gaussian likelihood has a higher Bayesian evidence than the generalised normal likelihood for the signal model $M_1$ with a Bayes factor of $\log K_{M_1, \Delta \mathcal{L}} = 1.53 \pm 0.39$. This is due to the Occams razor penalising the generalised normal likelihood for having an extra shape parameter $\beta_\mathrm{L}$ in its likelihood function that is not necessarily needed to model the Gaussian noise distribution of the dataset, hence, preferring the Gaussian likelihood function.
When we decrease the shape parameter $\beta_{\mathrm{D}} < 2$, the noise distribution turns to a heavy-tailed distribution, therefore, allowing more extreme noise realisations. For these cases e.g. $\beta_D \leq 1.2$, there is a clear preference ($\log K_{M_1, \Delta \mathcal{L}} < -3.50 \pm 0.41$) for the generalised normal likelihood over the Gaussian likelihood with the signal model having the highest Bayesian evidence of all models considered. This is due to the added flexibility of the generalised normal likelihood through its shape parameter $\beta_\mathrm{L}$. Similar behaviour can be seen when increasing the shape parameter i.e. concentrating most of the probability mass around the mean of the noise distribution. This likelihood preference has an impact on the sky-averaged 21-cm signal recovery which tends to be biased when using the Gaussian likelihood i.e. the posterior sky-averaged 21-cm signal amplitude is larger than expected. For example in the $\beta_\mathrm{D} =1$ case shown in Figure (\ref{fig:triangle_plot_beta1}), the posterior estimates of the amplitude (ground truth: $A_{21} = 155$ mK) using a Gaussian likelihood are $\bar A^*_{21} \approx 185 \pm 18$ mK whereas the generalised normal likelihood recovers the signal more accurately (here: $\bar A^*_{21} \approx 170 \pm 14$ mK).

For the no-signal model $M_2$, there is a tendency that the Gaussian likelihood is slightly more preferred than the generalised normal distribution for heavy-tailed noise ($\log K_{M_2, \Delta \mathcal{L}} \approx 1.90 \pm 0.40)$. This is due to the relatively weak sky-averaged 21-cm signal within the dataset. The no-signal model can use the foreground model to ``fit away'' the sky-averaged 21-cm signal and some of the extremer noise structures, hence, not requiring the additional flexibility of the generalised normal likelihood. 

We confirm this by regenerating the dataset with a larger sky-averaged 21-cm signal $A_{21} = 500$ mK. For these datasets and for the no-signal model $M_2$, the generalised normal likelihood is preferred over the Gaussian likelihood, as seen in Table \ref{tab:GenNormlogZRegenerated}. For these cases, the foreground and Gaussian likelihood function alone can not account for this larger structure, hence, requiring the additional flexibility of the generalised normal likelihood. We can see similar behaviour when we regenerate datasets where we increase the scale of the non-Gaussian noise to higher values $\sigma_{\mathrm{noise}}= 50$ mK. Here, the apparent preference of the Gaussian likelihood for the no-signal model vanishes, as the Bayes Factor is around zero.

\begin{table}
    \centering
    \begin{tabular}{lll}
\toprule
{}
Change & Noise parameter & $M_2, \mathrm{Gauss-GenNorm}$ \\
in Dataset & in Dataset  &    $\log K$   \\
\midrule
$A_{21} = 500$ mK & $\beta_{\mathrm{D}} = 1$  & $-2.90 \pm 0.38$\\
$A_{21} = 500$ mK & $\beta_{\mathrm{D}} = 1.2$  & $-3.00 \pm 0.38$\\
$A_{21} = 500$ mK & $\beta_{\mathrm{D}} = 1.5$  & $-5.11 \pm 0.36$ \\
$\sigma_{\mathrm{D}}= 50$ mK & $\beta_{\mathrm{D}} = 1$  & $-0.81 \pm 0.51$ \\
$\sigma_{\mathrm{D}}= 50$ mK & $\beta_{\mathrm{D}} = 1.2$  & $-0.50 \pm 0.37$ \\
$\sigma_{\mathrm{D}}= 50$ mK & $\beta_{\mathrm{D}} = 1.5$  & $+1.18 \pm 0.38$  \\
\bottomrule
\end{tabular}

    \caption{The Bayes factor for the no-signal model $M_2$ when using the Gaussian and generalised normal likelihood functions applied on a slightly modified dataset i.e. stronger sky-averaged 21-cm signal or larger noise. The noise parameter refers to the generalised normal noise distribution of the dataset.}
    \label{tab:GenNormlogZRegenerated}
\end{table}

\subsection{Student-t noise}
For the Student-t noise datasets, we see a tendency in favour of the Gaussian likelihood for the true signal model $M_1$ when the degree of freedom parameter $\nu$ is increasing (e.g. for $\nu = 30$ $\log K_{M_1, \Delta \mathcal{L}} = 1.59 \pm 0.40$). This is due to the Student-t distributions property to converge (in distribution) to a Gaussian distribution the more underlying samples one has. Therefore, Occam's razor penalises the generalised normal likelihood for its shape parameter and selects the Gaussian likelihood with the true signal model as the datasets best fitting choice. Moreover, the signal recovery is for both likelihoods almost identical with no signs of bias in the sky-averaged 21-cm signal posterior parameters as these Student-t noise distributions are similar to the Gaussian distribution, therefore modelled well with the likelihoods considered.

Consequently, when we decrease the degree of freedom parameter $\nu$, the generalised normal distribution will be preferred over the Gaussian likelihood model with Bayes factors of $\log K_{M_1, \Delta \mathcal{L}}  < -1.57 \pm 0.40$. However, we note for these heavy-tailed noise cases $\nu \leq 5$ , the no-signal model $M_2$ has a higher Bayesian evidence than the signal model $M_1$ with Bayes factors $\log K_{\Delta M } < -6$. This indicates that the signal-to-noise ratio is too low to confidently detect a sky-averaged 21-cm signal, therefore, preferring the simpler no-signal model. In terms of likelihood preference, the no-signal model has a tendency to prefer the simpler Gaussian likelihood function similar to the generalised normal noise case ($\log K_{M_2, \Delta \mathcal{L}} \approx 1.00 \pm 0.40$). Hence, the foreground model is used to ``fit away'' the sky-averaged 21-cm signal features and absorbing the structure into the noise, therefore, not needing the extra flexibility offered by the shape parameter of the generalised normal likelihood function.

\subsection{Cauchy noise}

For the Cauchy noise datasets, we conduct the inference with three different likelihoods. We use the Gaussian, generalised normal and Cauchy likelihood which is the ground truth likelihood function of the dataset. Here, the Gaussian likelihood has the lowest Bayesian evidence, irrespective of the scale $\gamma$ and model we consider. Only when decreasing the scale parameter $\gamma$ significantly, the Gaussian likelihoods perform gradually better, however still significantly unfavored with $\log K_{M_1, \Delta \mathcal{L}} < -140$.

In the signal model comparison, the true signal model $M_1$ is always preferred with $\log K_{\Delta M, \mathrm{Gauss}} > 2.7$, $\log K_{\Delta M, \mathrm{Gen. norm}} > 17$ and $\log K_{\Delta M, \mathrm{Cauchy}} > 15$. Furthermore, as expected, the Bayesian evidence is the highest for the Cauchy likelihood signal model, as it is the true noise model of the dataset. However, we note that the generalised normal likelihood performs almost equally well on the absolute log-evidence scale i.e. in the likelihood comparison, the Bayes factor is on a similar scale for the Cauchy and generalised normal likelihood function (e.g. for $\gamma = 0.025$ we have $\log K_{M_1, \mathrm{Gauss - Gen. norm}} =-340.35 \pm 0.39$ and $\log K_{M_1, \mathrm{Gauss - Cauchy}} = -356.00 \pm 0.37$) when compared to the Gaussian likelihood model. We note that these Bayes factors show a clear preference for the Cauchy likelihood when directly comparing it with the generalised normal likelihood function, however this Bayes factor difference is significantly smaller than the Gaussian likelihood comparison.

This Bayesian evidence difference between the likelihoods is noticed in the sky-averaged 21-cm signal recovery which is shown in Figure (\ref{fig:signalrecovery}). The generalised normal and Cauchy likelihood (ground truth) recovers the sky-averaged 21-cm signal well, however when using the Gaussian likelihood the sky-averaged 21-cm signal tends to have a noticeably larger amplitude i.e. the Gaussian likelihood fails to recover the true signal parameters. Thus, the generalised normal likelihood function can be considered as a good approximation of the true Cauchy likelihood function for sky-averaged 21-cm signal antenna temperature datasets containing Cauchy noise i.e. extreme noisy features.

\subsection{Radiometric noise}
For the radiometric noise datasets, we see a similar behaviour as the Cauchy noise case. The Gaussian likelihood always has the lowest Bayesian evidence for varying integration times $\tau$. For these datasets, the true radiometric likelihood has the highest Bayesian evidence. Similarly, as in the Cauchy noise datasets, the generalised normal likelihood performs reasonably well, hence, accounting for heteroscedastic noise structures through its shape parameter (e.g. for $\tau = 10^4 s$ we have $\log K_{M_1, \mathrm{Gauss - Gen. norm}} =-41.69 \pm 0.38$ and $\log K_{M_1, \mathrm{Gauss - Radiometric}} = -79.89 \pm 0.39$). This Bayes factor difference is more prominent for lower integration times. Moreover, this difference can be seen in the sky-averaged 21-cm signal recovery of Figure (\ref{fig:signalrecovery}) where the generalised normal and radiometric likelihood function recovers more tightly constrained posterior parameters in contrast to the homoscedastic Gaussian likelihood function that is more uncertain in the signal shape. 

In the signal comparison, we see a change in model preference at integration times of $\tau \sim 10^3$ s. For lower integration times, the signal-to-noise ratio is too low to make a confident sky-averaged 21-cm signal detection, therefore, preferring the no-signal model for these integration times e.g. $\log K_{\Delta M, \mathrm{Gen. norm}} \leq 0 \pm 0.42$ and $\log K_{\Delta M, \mathrm{Radiometric}} \leq 0 \pm 0.30$. Consequently, when we increase the integration time the radiometric noise got sufficiently suppressed such that the signal model will be preferred.

However, with the Gaussian likelihood the signal model $M_1$ is preferred again with Bayes Factors reaching up to $\log K_{\Delta M} = 37.11 \pm 0.39 $ if one keeps lowering the integration time. This is due to the models attempt to fit for a sky-averaged 21-cm signal imitated by the noise, therefore, incorrectly recovering the sky-averaged 21-cm signal parameters. We can mitigate this false sky-averaged 21-cm signal parameter recovery by changing the central frequency $f_{0,21}$ prior range. However, we do not see this behaviour for the generalised normal and radiometric likelihoods, as they correctly determine that these are noise structures, whereas the relatively constrained Gaussian likelihood failed to do so. Hence, we conclude that the generalized normal likelihood is a better approximation of the radiometric likelihood function if radiometric noise is present.

\begin{table*}[]
    \centering
    \caption{Bayesian evidences $\log \mathcal{Z}$ of all models and likelihoods considered for a simulated antenna temperature dataset with varying degrees and types of noise.}
    \label{tab:result_evidences}
\begin{tabular}{ccccc}
\toprule
{} $\textbf{Generalised normal noise }$&  Gauss $M_1, \log \mathcal{Z}$ &  Gen. norm $M_1, \log \mathcal{Z}$&  Gauss $M_2, \log \mathcal{Z}$ &  Gen. norm $M_2, \log \mathcal{Z}$ \\
\midrule
$\beta_D = 5.0$ &   $287.46 \pm 0.26$ &     $288.26\pm 0.28$ &  $251.13 \pm0.28$ &     $252.53 \pm 0.27$  \\
$\beta_D = 3.0$ &   $287.503 \pm0.28$ &     $286.10 \pm 0.27$&  $252.04 \pm0.29$ &     $251.74 \pm  0.28$\\
$\beta_D = 2.5$ &   $288.41 \pm0.27$&     $285.51\pm 0.27$&  $252.30\pm 0.28$ &     $250.77 \pm0.28$ \\
$\beta_D = 2.0$ &   $288.01 \pm0.28$ &     $286.482 \pm 0.27$&  $251.87 \pm0.28$&     $250.60\pm  0.27$\\
$\beta_D = 1.7$ &   $287.90 \pm0.28$&     $287.15 \pm 0.26$&  $253.23  \pm 0.27$ &     $251.60 \pm 0.27$ \\
$\beta_D = 1.5$ &   $288.54 \pm0.26$&     $288.29 \pm 0.28$ &  $253.18\pm 0.27$ &     $250.93 \pm 0.27$\\
$\beta_D = 1.2$ &   $288.32 \pm0.29$ &     $291.82 \pm 0.28$  &  $253.69 \pm 0.28$  &     $251.37 \pm  0.27$\\
$\beta_D = 1.0$ &   $289.57 \pm0.28$&     $297.15 \pm0.28$ &  $254.15 \pm0.29$  &     $252.03\pm 0.26$\\
\end{tabular}

\begin{tabular}{ccccc}
\toprule
{} $\textbf{Student-t noise}$&  Gauss, $M_1, \log \mathcal{Z}$ &  Gen. norm $M_1, \log \mathcal{Z}$ &  Gauss $M_2, \log \mathcal{Z}$ &  Gen. norm $M_2, \log \mathcal{Z}$ \\
\midrule
$\nu = 30$ &   $297.86 \pm0.29$ &     $296.26 \pm0.28$ &  $256.82 \pm0.29$&     $255.35\pm 0.27$\\
$\nu = 20$ &   $294.83 \pm0.29$ &     $295.91 \pm0.29$&  $257.11 \pm0.28$&     $255.99 \pm0.27$\\
$\nu = 15$ &   $293.40 \pm0.29$&     $293.88 \pm0.29$&  $256.76 \pm0.28$&     $254.53\pm 0.28$\\
$\nu = 10$ &   $287.28 \pm0.29$&     $288.86 \pm0.29$&  $256.29 \pm0.29$&     $255.55 \pm0.27$\\
$\nu = 5$ &   $246.99 \pm0.28$&     $248.87 \pm0.29$&  $255.96\pm 0.29$&     $255.86 \pm0.29$\\
$\nu = 4$ &   $239.76 \pm0.27$&     $246.30 \pm0.29$&  $256.01 \pm0.28$&     $255.81 \pm0.29$\\
$\nu = 3$ &   $221.29 \pm0.28$&     $230.73 \pm0.28$&  $256.45 \pm0.28$&     $255.43 \pm0.28$\\
$\nu = 2$ &   $194.14 \pm0.27$&     $224.73 \pm0.26$&  $257.10 \pm0.30$&     $254.95 \pm0.28$ \\

\end{tabular}

\begin{tabular}{ccccccc}
\toprule
{} $\textbf{Cauchy noise}$ &  Gauss $M_1, \log \mathcal{Z}$ &  Gen. norm $M_1, \log \mathcal{Z}$ &  Cauchy $M_1, \log \mathcal{Z}$ &  Gauss $M_2, \log \mathcal{Z}$ &  Gen. norm $M_2, \log \mathcal{Z}$ &  Cauchy $M_2, \log \mathcal{Z}$ \\
\midrule
$\gamma = 0.03$ &  $-385.19 \pm0.26$&      $90.98 \pm0.29$&     $110.43 \pm0.27$& $-392.70 \pm 0.29$&      $73.96 \pm0.29$&      $95.01 \pm0.27$\\
$\gamma = 0.025$ &  $-219.19 \pm0.26$&     $121.15 \pm0.30$&     $136.81 \pm0.26$& $-225.72 \pm0.28$&      $99.11 \pm0.31$&     $115.21 \pm0.29$\\
$\gamma = 0.02$ &   $-83.17 \pm0.27$&     $155.39 \pm0.29$&     $167.82 \pm0.28$&  $-88.21 \pm0.28$&     $129.44 \pm0.30$&     $139.24 \pm0.27$ \\
$\gamma = 0.015$ &    $23.35 \pm0.27$ &     $197.08 \pm0.31$&     $207.64 \pm0.27$ &   $19.23 \pm0.27$&     $158.52 \pm0.30$&     $167.14 \pm0.27$\\
$\gamma = 0.01$ &   $102.48 \pm0.24$&     $253.97  \pm0.30$&     $263.31 \pm0.29$&   $99.69 \pm0.26$&     $193.33 \pm029$&     $200.85 \pm0.27$\\
$\gamma = 0.005$ &   $199.89 \pm0.26$&     $344.75 \pm0.31$&     $352.83 \pm0.31$&  $188.10 \pm0.28$&     $242.32 \pm0.30$&     $245.56 \pm0.29$\\

\end{tabular}

\begin{tabular}{ccccccc}
\toprule
{} $\textbf{Radiometric noise}$&    Gauss $M_1, \log \mathcal{Z}$ &  Gen. norm $M_1, \log \mathcal{Z}$ &  Radiometric $M_1, \log \mathcal{Z}$ &     Gauss $M_2, \log \mathcal{Z}$ &  Gen. norm $M_2, \log \mathcal{Z}$ &  Radiometric $M_2, \log \mathcal{Z}$ \\
\midrule
$\tau = 10^5 $ s &     $315.73 \pm0.28$&     $353.55 \pm0.32$&     $392.05 \pm0.28$ &     $278.62 \pm0.28$&     $279.97 \pm0.29$&     $309.10 \pm0.26$\\
$\tau = 10^4 $ s &     $155.56 \pm0.26$&     $197.25 \pm0.28$&     $235.45 \pm0.25$&     $153.56 \pm0.27$&     $179.63 \pm0.27$&     $220.37 \pm0.25$\\
$\tau = 10^3 $ s &    $-131.25 \pm0.26$&      $11.52 \pm0.27$&      $74.20 \pm0.22$&    $-131.22 \pm0.28$&       $6.85 \pm0.28$&      $72.76 \pm0.23$\\
$\tau = 10^2 $ s &   $-2792.04 \pm0.28$&    $-341.69 \pm0.28$&     $-90.55 \pm0.20$&   $-2793.34 \pm0.27$&   $-342.62 \pm0.30$&     $-90.56 \pm0.22$\\
$\tau = 10^1 $ s&  $-29840.63 \pm0.30$&    $-820.28 \pm0.29$&    $-257.50 \pm0.17$&  $-29873.58 \pm0.29$&    $-818.71 \pm0.33$&    $-257.56 \pm0.19$\\
$\tau = 10^0 $ s & $-304196.93 \pm0.32$&   $-1325.48 \pm0.36$&    $-427.23 \pm0.16$& $-304388.36 \pm0.30$&   $-1326.45 \pm0.35$&    $-426.57 \pm0.17$\\

\bottomrule
\end{tabular}
\end{table*}

\section{CONCLUSION}
\label{sec:Summary}
For Bayesian inference, the necessity for an explicit likelihood function poses a challenging task for cosmological state-of-the-art data analysis. Usually, it is easier to simulate a dataset for a physical process based on some input parameters than to identify the often latent and hierarchical statistical properties in each model component and solving the inverse problem. Therefore, these complex noise structures can make the likelihood intractable and a more simplified approximation such as a Gaussian likelihood is frequently used. The task of finding an approximation of the ground truth likelihood function is known as likelihood-free inference (LFI) with often complex and involved algorithms executed to achieve this. However, we demonstrated that LFI can also be achieved through a simple Bayesian evidence-driven likelihood analysis in which we show that the Gaussian likelihood function is not always the best choice as an approximation of the ground truth likelihood function of the dataset. Although we note, that in `true' LFI the non-parametric likelihood approximations are much more expressive than those considered here, nonetheless they command much larger Occam penalties given by the virtue of their expressiveness. This forms the subject of ongoing work.

We used a physically motivated simulator to generate antenna temperature datasets where we injected noise of various complexity and scales. For each dataset, we injected one of the following noise structures: Gaussian noise, generalised normal noise, Student-t noise, Cauchy noise or radiometric noise. As each of these noise distributions has a varying amount of free parameters, we varied the shape parameter, if the distribution has a shape parameter, otherwise we varied its scale parameter.

After dataset creation, we defined two models: the true signal model $M_1$ of the dataset containing the foreground, the sky-averaged 21-cm signal and the noise contribution and the no signal model $M_2$ which does not contain any sky-averaged 21-cm signal component.

We modelled the foreground by dividing the sky into $N_{\mathrm{reg}} = 14$ regions of equal spectral index and the sky-averaged 21-cm signal through a Gaussian-shaped absorption feature and the noise contribution through the likelihood function. As we injected noise with various structures, we investigated the following likelihood functions: the Gaussian, generalised normal, Cauchy and the radiometric likelihood. 

By providing priors for our model parameters, we used \texttt{PolyChord}, a nested sampling-based algorithm, to numerically compute the Bayesian evidence of the model and likelihood. Afterward, we selected the preferred likelihood through the Bayesian evidence for each model and dataset.

We found that if the dataset contains Gaussian or Gaussian-like noise (e.g. Student-t with high $\nu$), the Gaussian likelihood function has the highest Bayesian evidence for the true signal model $M_1$ of the dataset. However, if the noise is non-Gaussian/heavy-tailed e.g. Cauchy or radiometric noise, the Gaussian likelihood has always a lower Bayesian evidence with a tendency for biased signal parameter recovery than the generalised normal and ground truth likelihood functions for the datasets studied. Consequently, for these noise structures, the respective ground truth likelihoods e.g. Cauchy likelihood for Cauchy noise or the radiometric likelihood for radiometric noise has the highest Bayesian evidence. However, the generalised normal likelihood is only marginally worse on Bayesian evidence scale with comparable sky-averaged 21-cm signal recovery than the ground truth likelihood functions. This is due to the flexibility added by the shape parameter of the generalised normal distribution to account for extremely noisy features.

To summarise, this analysis demonstrated the influence of varying likelihoods on sky-averaged 21-cm signal recovery and it showed that one can get biased posterior parameter results such as amplitude enhancement or more uncertain shape parameters by using the oversimplified Gaussian likelihood function for inference. Hence, only if one is certain that the noise structure of the antenna temperature dataset is Gaussian one should use the Gaussian likelihood function for inference.  Moreover, in the general case, if one is certain about the precise noise structure, one should always use its ground truth likelihood function. However, when one is uncertain about the noise structure, which in principle holds true for every complex physical process such as the challenging task of sky-averaged 21-cm signal recovery, one should use the generalised normal likelihood as a preferred first approximation for the true unknown likelihood function and expand on it with further analysis. The generalised normal likelihood offers the flexibility to account for more complex unknown noise structure and it generalises the Gaussian distribution by including it as a special case. Hence, this form of preference of the generalised normal likelihood selected by the Bayesian evidence is a simple way of finding a good approximation of the ground truth likelihood with wider applications in the likelihood-free inference paradigm by quantifying how well the likelihood estimate found through an algorithmic method is.

\begin{acknowledgement}
This work was performed using resources provided by the Cambridge Service for Data Driven Discovery (CSD3) operated by the University of Cambridge Research Computing Service, provided by Dell EMC and Intel using Tier-2 funding from the Engineering and Physical Sciences Research Council (capital grant EP/T022159/1), and DiRAC funding from the Science and Technology Facilities Council. KHS would like to thank Dominic Anstey for providing the core Bayesian data analysis pipeline used in REACH and the Hans Werthén foundation of the Royal Swedish academy of Engineering Sciences. WH is funded by a Gonville \& Caius College Research Fellowship and Royal Society University Research Fellowship. EdLA is funded by the Science and Technologies Facilities Council Ernest Rutherford Fellowship.
\end{acknowledgement}

\section*{Data Availability}
The nested sampling posterior samples are available on Zenodo at: \href{https://doi.org/10.5281/zenodo.6473727}{https://doi.org/10.5281/zenodo.6473727}


\bibliography{references}

\begin{thebibliography}{}
\expandafter\ifx\csname natexlab\endcsname\relax\def\natexlab#1{#1}\fi

\bibitem[{Acedo {et~al.}(2015)Acedo, Razavi-Ghods, Troop, Drought, \&
  Faulkner}]{acedo_skala_2015}
Acedo, E. d.~L., Razavi-Ghods, N., Troop, N., Drought, N., \& Faulkner, A.~J.
  2015, Experimental Astronomy, 39, 567, arXiv:1512.01453 [astro-ph]

\bibitem[{Alsing {et~al.}(2019)Alsing, Charnock, Feeney, \&
  Wandelt}]{alsing_fast_2019}
Alsing, J., Charnock, T., Feeney, S., \& Wandelt, B. 2019, Monthly Notices of
  the Royal Astronomical Society, 488, 4440

\bibitem[{Alsing {et~al.}(2018)Alsing, Wandelt, \&
  Feeney}]{alsing_massive_2018}
Alsing, J., Wandelt, B., \& Feeney, S. 2018, Monthly Notices of the Royal
  Astronomical Society, 477, 2874, arXiv: 1801.01497

\bibitem[{Anstey {et~al.}(2021{\natexlab{a}})Anstey, Cumner, de~Lera~Acedo, \&
  Handley}]{anstey_informing_2021}
Anstey, D., Cumner, J., de~Lera~Acedo, E., \& Handley, W. 2021{\natexlab{a}},
  Monthly Notices of the Royal Astronomical Society, 509, 4679

\bibitem[{Anstey {et~al.}(2021{\natexlab{b}})Anstey, de Lera Acedo, \&
  Handley}]{anstey_general_2021}
Anstey, D., de Lera Acedo, E., \& Handley, W. 2021{\natexlab{b}}, Monthly
  Notices of the Royal Astronomical Society, 506, 2041

\bibitem[{Barkana(2018)}]{barkana_possible_2018}
Barkana, R. 2018, Nature, 555, 71, number: 7694 Publisher: Nature Publishing
  Group

\bibitem[{Barkana {et~al.}(2018)Barkana, Outmezguine, Redigolo, \&
  Volansky}]{barkana_signs_2018}
Barkana, R., Outmezguine, N.~J., Redigolo, D., \& Volansky, T. 2018, Physical
  Review D, 98, 103005, arXiv: 1803.03091

\bibitem[{Bevins {et~al.}(2021)Bevins, Handley, Fialkov, de Lera Acedo,
  Greenhill, \& Price}]{bevins_maxsmooth_2021}
Bevins, H. T.~J., Handley, W.~J., Fialkov, A., {et~al.} 2021, Monthly Notices
  of the Royal Astronomical Society, 502, 4405

\bibitem[{Bowman {et~al.}(2018{\natexlab{a}})Bowman, Rogers, Monsalve, Mozdzen,
  \& Mahesh}]{bowman_absorption_2018}
Bowman, J.~D., Rogers, A. E.~E., Monsalve, R.~A., Mozdzen, T.~J., \& Mahesh, N.
  2018{\natexlab{a}}, Nature, 555, 67, number: 7694 Publisher: Nature
  Publishing Group

\bibitem[{Bowman {et~al.}(2018{\natexlab{b}})Bowman, Rogers, Monsalve, Mozdzen,
  \& Mahesh}]{bowman_reply_2018}
---. 2018{\natexlab{b}}, Nature, 564, E35, number: 7736 Publisher: Nature
  Publishing Group

\bibitem[{Bradley {et~al.}(2019)Bradley, Tauscher, Rapetti, \&
  Burns}]{bradley_ground_2019}
Bradley, R.~F., Tauscher, K., Rapetti, D., \& Burns, J.~O. 2019, The
  Astrophysical Journal, 874, 153, arXiv: 1810.09015

\bibitem[{Cohen {et~al.}(2017)Cohen, Fialkov, Barkana, \&
  Lotem}]{cohen_charting_2017}
Cohen, A., Fialkov, A., Barkana, R., \& Lotem, M. 2017, Monthly Notices of the
  Royal Astronomical Society, 472, 1915, arXiv: 1609.02312

\bibitem[{Cranmer {et~al.}(2020)Cranmer, Brehmer, \&
  Louppe}]{cranmer_frontier_2020}
Cranmer, K., Brehmer, J., \& Louppe, G. 2020, Proceedings of the National
  Academy of Sciences, 117, 30055, publisher: National Academy of Sciences
  Section: Colloquium Paper

\bibitem[{de~Lera~Acedo {et~al.}(2022)de~Lera~Acedo, de~Villiers, Razavi-Ghods,
  Handley, Fialkov, Magro, Anstey, Bevins, Chiello, Cumner, Josaitis, Roque,
  Sims, Scheutwinkel, Alexander, Bernardi, Carey, Cavillot, Croukamp, Ely,
  Gessey-Jones, Gueuning, Hills, Kulkarni, Maiolino, Meerburg, Mittal,
  Pritchard, Puchwein, Saxena, Shen, Smirnov, Spinelli, \&
  Zarb-Adami}]{de_lera_acedo_reach_2022}
de~Lera~Acedo, E., de~Villiers, D. I.~L., Razavi-Ghods, N., {et~al.} 2022,
  Nature Astronomy, 1, publisher: Nature Publishing Group

\bibitem[{De~Oliveira-Costa {et~al.}(2008)De~Oliveira-Costa, Tegmark, Gaensler,
  Jonas, Landecker, \& Reich}]{de_oliveira-costa_model_2008}
De~Oliveira-Costa, A., Tegmark, M., Gaensler, B.~M., {et~al.} 2008, Monthly
  Notices of the Royal Astronomical Society, 388, 247

\bibitem[{Dyson(1965)}]{dyson_characteristics_1965}
Dyson, J. 1965, IEEE Transactions on Antennas and Propagation, 13, 488,
  conference Name: IEEE Transactions on Antennas and Propagation

\bibitem[{Elsherbeni(2014)}]{elsherbeni_antenna_2014}
Elsherbeni, A. Z.~a. 2014, Antenna analysis and design using {FEKO}
  electromagnetic simulation software (Edison, NJ : SciTech Publishing, an
  imprint of the IET, [2014] ©2014)

\bibitem[{Feroz \& Hobson(2008)}]{feroz_multimodal_2008}
Feroz, F., \& Hobson, M.~P. 2008, Monthly Notices of the Royal Astronomical
  Society, 384, 449

\bibitem[{Feroz {et~al.}(2009)Feroz, Hobson, \& Bridges}]{feroz_multinest_2009}
Feroz, F., Hobson, M.~P., \& Bridges, M. 2009, Monthly Notices of the Royal
  Astronomical Society, 398, 1601

\bibitem[{Fialkov \& Barkana(2019)}]{fialkov_signature_2019}
Fialkov, A., \& Barkana, R. 2019, Monthly Notices of the Royal Astronomical
  Society, 486, 1763, arXiv: 1902.02438

\bibitem[{Furlanetto {et~al.}(2006)Furlanetto, Oh, \&
  Briggs}]{furlanetto_cosmology_2006}
Furlanetto, S., Oh, S.~P., \& Briggs, F. 2006, Physics Reports, 433, 181,
  arXiv: astro-ph/0608032

\bibitem[{Handley {et~al.}(2015{\natexlab{a}})Handley, Hobson, \&
  Lasenby}]{handley_polychord_2015}
Handley, W.~J., Hobson, M.~P., \& Lasenby, A.~N. 2015{\natexlab{a}}, Monthly
  Notices of the Royal Astronomical Society: Letters, 450, L61, arXiv:
  1502.01856

\bibitem[{Handley {et~al.}(2015{\natexlab{b}})Handley, Hobson, \&
  Lasenby}]{handley_polychord_2015-1}
---. 2015{\natexlab{b}}, Monthly Notices of the Royal Astronomical Society,
  453, 4385, arXiv: 1506.00171

\bibitem[{Hills {et~al.}(2018)Hills, Kulkarni, Meerburg, \&
  Puchwein}]{hills_concerns_2018}
Hills, R., Kulkarni, G., Meerburg, P.~D., \& Puchwein, E. 2018, Nature, 564,
  E32, arXiv: 1805.01421

\bibitem[{Jana {et~al.}(2019)Jana, Nath, \& Biermann}]{jana_radio_2019}
Jana, R., Nath, B.~B., \& Biermann, P.~L. 2019, Monthly Notices of the Royal
  Astronomical Society, 483, 5329, arXiv: 1812.07404

\bibitem[{Jeffrey {et~al.}(2020)Jeffrey, Alsing, \&
  Lanusse}]{jeffrey_likelihood-free_2020}
Jeffrey, N., Alsing, J., \& Lanusse, F. 2020, Monthly Notices of the Royal
  Astronomical Society, 501, 954, arXiv: 2009.08459

\bibitem[{Kraus {et~al.}(1986)Kraus, Tiuri, Räisänen, \&
  Carr}]{kraus_radio_1986}
Kraus, J.~D., Tiuri, M., Räisänen, A.~V., \& Carr, T.~D. 1986, Radio
  {Astronomy} (Cygnus-Quasar Books), google-Books-ID: KtVFAQAAIAAJ

\bibitem[{Liu {et~al.}(2013)Liu, Pritchard, Tegmark, \& Loeb}]{liu_global_2013}
Liu, A., Pritchard, J.~R., Tegmark, M., \& Loeb, A. 2013, Physical Review D,
  87, 043002, arXiv: 1211.3743

\bibitem[{MacKay(2003)}]{mackay_information_2003}
MacKay, D. J.~C. 2003, Information {Theory}, {Inference}, and {Learning}
  {Algorithms}

\bibitem[{Marin {et~al.}(2011)Marin, Pudlo, Robert, \&
  Ryder}]{marin_approximate_2011}
Marin, J.-M., Pudlo, P., Robert, C.~P., \& Ryder, R. 2011, Approximate
  {Bayesian} {Computational} methods, Tech. rep., publication Title: arXiv
  e-prints ADS Bibcode: 2011arXiv1101.0955M Type: article

\bibitem[{Mirocha \& Furlanetto(2019)}]{mirocha_what_2019}
Mirocha, J., \& Furlanetto, S.~R. 2019, Monthly Notices of the Royal
  Astronomical Society, 483, 1980, arXiv: 1803.03272

\bibitem[{Mittal \& Kulkarni(2022)}]{mittal_implications_2022}
Mittal, S., \& Kulkarni, G. 2022, arXiv:2203.07733 [astro-ph], arXiv:
  2203.07733

\bibitem[{Muñoz \& Loeb(2018)}]{munoz_insights_2018}
Muñoz, J.~B., \& Loeb, A. 2018, Nature, 557, 684, arXiv: 1802.10094

\bibitem[{Papamakarios \& Murray(2016)}]{papamakarios_fast_2016}
Papamakarios, G., \& Murray, I. 2016, Fast $\epsilon$-free {Inference} of
  {Simulation} {Models} with {Bayesian} {Conditional} {Density} {Estimation},
  Tech. rep., publication Title: arXiv e-prints ADS Bibcode:
  2016arXiv160506376P Type: article

\bibitem[{Papamakarios {et~al.}(2021)Papamakarios, Nalisnick, Rezende, Mohamed,
  \& Lakshminarayanan}]{papamakarios_normalizing_2021}
Papamakarios, G., Nalisnick, E., Rezende, D.~J., Mohamed, S., \&
  Lakshminarayanan, B. 2021, Journal of Machine Learning Research, 22, 1

\bibitem[{Petrosyan \& Handley(2022)}]{petrosyan_supernest_2022}
Petrosyan, A., \& Handley, W.~J. 2022, {SuperNest}: accelerated nested sampling
  applied to astrophysics and cosmology, arXiv:2212.01760 [astro-ph,
  physics:physics], doi:10.48550/arXiv.2212.01760

\bibitem[{Price {et~al.}(2018)Price, Greenhill, Fialkov, Bernardi, Garsden,
  Barsdell, Kocz, Anderson, Bourke, Craig, Dexter, Dowell, Eastwood, Eftekhari,
  Ellingson, Hallinan, Hartman, Kimberk, Lazio, Leiker, MacMahon, Monroe,
  Schinzel, Taylor, Tong, Werthimer, \& Woody}]{price_design_2018}
Price, D.~C., Greenhill, L.~J., Fialkov, A., {et~al.} 2018, Monthly Notices of
  the Royal Astronomical Society, 478, 4193

\bibitem[{Pritchard \& Loeb(2008)}]{pritchard_evolution_2008}
Pritchard, J.~R., \& Loeb, A. 2008, Physical Review D, 78, 103511, arXiv:
  0802.2102

\bibitem[{Roque {et~al.}(2021)Roque, Handley, \&
  Razavi-Ghods}]{roque_bayesian_2021}
Roque, I. L.~V., Handley, W.~J., \& Razavi-Ghods, N. 2021, Monthly Notices of
  the Royal Astronomical Society, 505, 2638, arXiv: 2011.14052

\bibitem[{Scheutwinkel {et~al.}(2022)Scheutwinkel, Acedo, \&
  Handley}]{scheutwinkel_bayesian_2022}
Scheutwinkel, K.~H., Acedo, E. d.~L., \& Handley, W. 2022, Publications of the
  Astronomical Society of Australia, 39, e052, publisher: Cambridge University
  Press

\bibitem[{Sims \& Pober(2020)}]{sims_testing_2020}
Sims, P.~H., \& Pober, J.~C. 2020, Monthly Notices of the Royal Astronomical
  Society, 492, 22, arXiv: 1910.03165

\bibitem[{Singh \& Subrahmanyan(2019)}]{singh_redshifted_2019}
Singh, S., \& Subrahmanyan, R. 2019, The Astrophysical Journal, 880, 26, arXiv:
  1903.04540

\bibitem[{Singh {et~al.}(2018)Singh, Subrahmanyan, Shankar, Rao, Girish,
  Raghunathan, Somashekar, \& Srivani}]{singh_saras_2018-1}
Singh, S., Subrahmanyan, R., Shankar, N.~U., {et~al.} 2018, Experimental
  Astronomy, 45, 269

\bibitem[{Singh {et~al.}(2021)Singh, T., Subrahmanyan, Shankar, Girish,
  Raghunathan, Somashekar, Srivani, \& Rao}]{singh_detection_2021}
Singh, S., T., J.~N., Subrahmanyan, R., {et~al.} 2021, arXiv:2112.06778
  [astro-ph], arXiv: 2112.06778

\bibitem[{Sivia \& Skilling(2006)}]{sivia_data_2006}
Sivia, D.~S., \& Skilling, J. 2006, Data analysis: a {Bayesian} tutorial, 2nd
  edn., Oxford science publications (Oxford, England: Oxford University Press)

\bibitem[{Skilling(2006)}]{skilling_nested_2006}
Skilling, J. 2006, Bayesian Analysis, 1, 833, publisher: International Society
  for Bayesian Analysis

\bibitem[{Zhao {et~al.}(2022{\natexlab{a}})Zhao, Mao, Cheng, \&
  Wandelt}]{zhao_simulation-based_2022}
Zhao, X., Mao, Y., Cheng, C., \& Wandelt, B.~D. 2022{\natexlab{a}}, The
  Astrophysical Journal, 926, 151, publisher: American Astronomical Society

\bibitem[{Zhao {et~al.}(2022{\natexlab{b}})Zhao, Mao, \&
  Wandelt}]{zhao_implicit_2022}
Zhao, X., Mao, Y., \& Wandelt, B.~D. 2022{\natexlab{b}}, arXiv:2203.15734
  [astro-ph], arXiv: 2203.15734

\end{thebibliography}

\end{document}